\documentclass[twocolumn]{aastex6}

\usepackage{hyperref}
\PassOptionsToPackage{pdfpagelabels=false, pageanchor=false, hyperfootnotes=false, plainpages=false,hypertexnames=false}{hyperref} 

\hypersetup{colorlinks,linkcolor=blue,citecolor=blue,urlcolor=blue}

\usepackage{newtxtext,newtxmath}

\usepackage{graphicx}	
\usepackage{amsmath}	
\usepackage{amssymb}	

\RequirePackage{rotating}

\usepackage[T1]{fontenc}
\usepackage{ae,aecompl}

\usepackage{textcomp}
\usepackage{gensymb}
\usepackage{etoolbox}
\usepackage{xspace}
\usepackage{paralist}
\usepackage{blindtext}
\usepackage[english]{babel}

\newcommand{\vRR}{\frac{\partial{<V_R>}}{\partial{R}}}

\newcommand{\pc}{\,\rm{pc}\xspace}
\newcommand{\kpc}{\,\rm{kpc}\xspace}
\newcommand{\kms}{\,\rm{km\,s^{-1}}\xspace}
\newcommand{\kmskpc}{\,\rm{km\,s^{-1}\,kpc^{-1}}\xspace}
\mathchardef\mhyphen="2D

\newcommand{\gaia}{\textit{Gaia}\xspace}
\newcommand{\tgas}{TGAS\xspace}
\newcommand{\rave}{RAVE\xspace}
\newcommand{\lamost}{LAMOST\xspace}

\newcommand{\Gyr}{\,\rm{Gyr}}
\newcommand{\Myr}{\,\rm{Myr}}

\newtoggle{comment}
\toggletrue{comment}

\received{}
\revised{}
\accepted{}

\shorttitle{Hercules stream}
\shortauthors{P\'erez-Villegas et al.}

\begin{document}


\title{Revisiting the Tale of Hercules: how stars orbiting the Lagrange points visit the Sun}



\author{Angeles P\'erez-Villegas\altaffilmark{1}, Matthieu Portail, Christopher Wegg and Ortwin Gerhard}
\affil{Max-Planck-Institut f\"ur Extraterrestrische Physik, Gie\ss enbachstra\ss e, D-85741 Garching, Germany}

\altaffiltext{1}{mperez@mpe.mpg.de}




\begin{abstract}
We propose a novel explanation for the Hercules stream consistent with recent measurements of the extent and pattern speed of the Galactic bar. We have adapted a made-to-measure dynamical model tailored for the Milky Way to investigate the kinematics of
the solar neighborhood (SNd). The model matches the 3D density of the red clump giant stars (RCGs) in the bulge and bar as well as stellar kinematics in the inner Galaxy, with a pattern speed of $39\kmskpc$. Cross-matching this model
with the $Gaia$ DR1 TGAS data combined with \rave and \lamost radial velocities, we find that the model naturally
predicts a bimodality in the {\it U--V}-velocity distribution for nearby stars which is in good agreement with the
Hercules stream.  In the model, the Hercules stream is made of stars orbiting the Lagrange points of the bar
which move outward from the bar's corotation radius to visit the SNd. 
While the model is not yet a quantitative fit of the velocity distribution, the new picture naturally predicts that the Hercules stream is more prominent inward from the Sun and nearly absent only a few $100 \pc$ outward of the Sun, and plausibly explains that Hercules is prominent in old and metal-rich stars.
\end{abstract}




\section{Introduction} \label{sec:intro}

{\it Hipparcos} data clearly revealed the rich substructure in the velocity distribution of solar neighborhood
(SNd) stars \citep{Dehnen1998}. Previously, \citet{Eggen1996} had related this structure with moving groups of stars that
retain the kinematic signature of their birth places. However, the kinematically identified groups also
contain old and late-type stars \citep{Dehnen1998, Antoja2008}, and some may be related to orbit structures
caused by the Galactic bar or spiral arms \citep{Dehnen2000, Quillen2005}.

One of the most studied kinematic group is known as the Hercules stream (or $U$-anomaly) identified by \citet{Eggen58}. The Hercules stream is an excess of stars with negative $U$ velocities (away from the Galactic center, GC) that also move slower than the Sun's velocity by $V\sim-50\kms$. \citet{Dehnen1998} used main-sequence stars from {\it Hipparcos} and separated them in $B-V$, showing that Hercules is prominent at $B-V>0.6$. Therefore, this stream is likely to have a dynamical origin both due to the stars being older than $\sim 4$ Gyr \citep{Antoja2008} and because they have a wide range of metallicities \citep{Raboud1998,Bensby2007}. 

A mechanism to explain the dynamical origin of the Hercules stream was proposed by \citet{Dehnen2000}, who suggested that these stars reach the SNd on orbits due to the Outer Lindblad resonance (OLR) of the Galactic bar. In this scenario, the OLR needs to be placed close to the Sun, implying a pattern speed of the bar of $\sim$$1.85$ times the local circular frequency \citep{Antoja2014,Monari2017}.

This model for the Hercules stream fits well with a short and fast-rotating bar, whose extension is between $3.0$ and $3.5\kpc$ \citep{Binney1997,Bissantz2002} with a pattern speed in the range of $50-65\kmskpc$ \citep{Englmaier1999,Fux1999}. However, recent measurements show that the Galactic long bar extends to $5.0\pm0.2\kpc$ from the Galactic center \citep{Wegg2015}, and from new dynamical modeling of the stellar kinematics of the inner Galaxy, the pattern speed is $39\pm3.5\kmskpc$ \citep{Portail2017}. Then, the OLR is placed at $\sim10.5\kpc$, too far outside the SNd for the Hercules stream to be explained by OLR orbits. 

In this Letter, we revisit the origin of the Hercules stream and show that these stars in the SNd can be
naturally explained in the framework of such a slow bar. In this interpretation, the Hercules stream is made of stars orbiting the Lagrange points
of the Galactic bar that reach the SNd while moving outward from the corotation (CR) region.

This work is organized as follows. In \autoref{section:modelAndData}, we describe our made-to-measure dynamical model of the Galaxy and the data we employed. In \autoref{section:hercules}, we show that the Hercules stream naturally arises in this model, discuss which kind of orbits cause the secondary peak in the SNd velocity distribution, and how the prominence of the Hercules stream changes with Galactocentric distance and metallicity. We summarize and discuss our results in \autoref{section:summary}.


\section{Model and data}
\label{section:modelAndData}

\subsection{Dynamical model of the Galaxy}
\label{section:model}

Recently, \citet{Portail2017} presented the first non-parametric dynamical model of the entire bar region. Using the Made-to-Measure (M2M) method \citep{DeLorenzi2007, Portail2015a}, they adapted the particle weights of a self-consistent $N$-body model of the Galaxy in order to match data from numerous surveys. Their model successfully reproduces the 3D density of RCGs in the bulge from \citet{Wegg2013}; the magnitude distributions of RCGs across the bar region from the combination of the VVV, UKIDSS, and 2MASS surveys \citep{Wegg2015}; stellar kinematics from the BRAVA \citep{Kunder2012}, OGLE-II \citep{Sumi2004}, and ARGOS \citep{Freeman2013, Ness2013b} surveys; and the outer rotation curve up to $10\kpc$ \citep{Sofue2009}.

A few adjustments to the fitted and phase-mixed model of \citet{Portail2017} are required in order to study the influence of the galactic bar on the SNd. First, we increase the resolution of the model by a factor of 10, using a variant of the resampling algorithm presented in \citet{Dehnen2009} because originally only $\sim650$ particles lay within 600 pc from the Sun. Second, the radial dispersion of the outer disk unconstrained in \citet{Portail2017} and is found after M2M fitting to be $\sigma_U=44\kms$, larger than the observed $\sim36\kms$ (see \autoref{section:data}). Large radial dispersion in the disk is the result of orbits with large epicycle motions around their guiding centers. We thus cool the outer disk by evolving it for $\sim$$1\Gyr$, during which we apply a damping term in the radial equation of motion for particles with guiding radius $>6.5\kpc$. The additional radial drag force used is $\dot{{\rm v}}_R=-2\,\lambda\,{\rm v}_R\,$, where $\lambda=2.5\Gyr^{-1}$ is the timescale of the disk cooling, and ${\rm v}_R$ the radial velocity of the particle. Once the disk is cooled, we reproduce the M2M fitting of \citet{Portail2017}, resulting in a nearly identical bar model, but with a high-resolution disk and a local radial dispersion of $35.5\kms$, in good agreement with the data.

\begin{figure}
  \centering
  \includegraphics[width=\columnwidth]{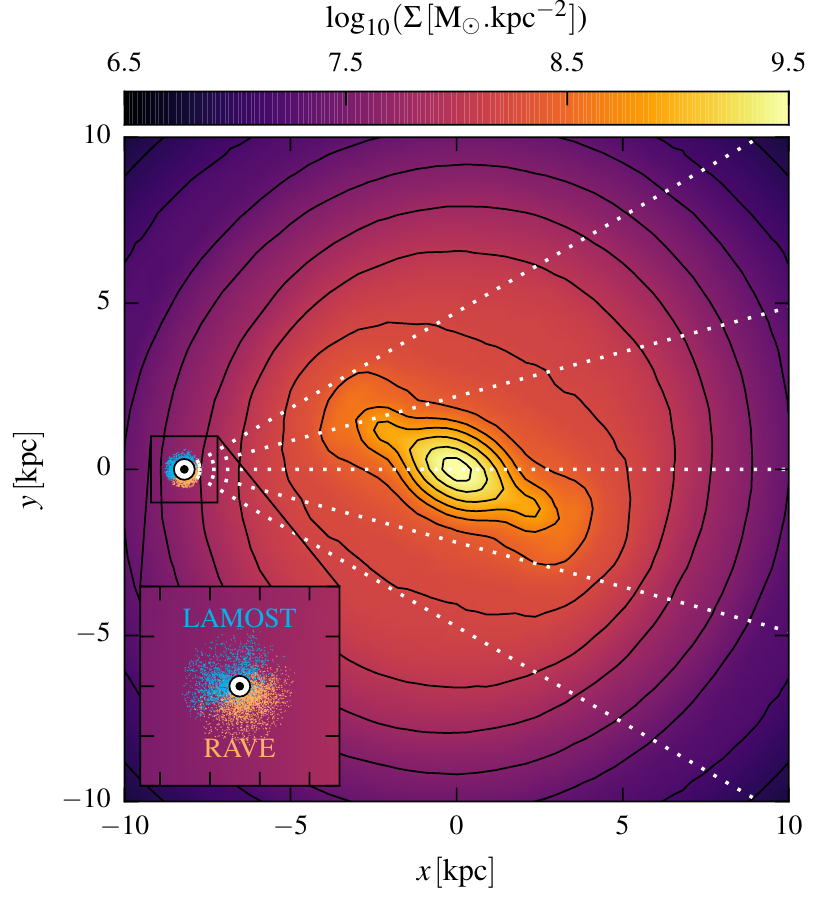}\\
  \caption{Face-on projection of the dynamical model of the galactic bar based on the work of \citet{Portail2017}. Sightlines from the Sun (large dot) are indicated for Galactic longitudes of $l=-30\degree,-15\degree,0\degree,+15\degree$, and $+30\degree$. The insert shows a zoom of the SNd where blue and yellow dots indicate the coverage of the TGAS+RAVE and TGAS+LAMOST data, respectively.}
  \label{fig:model}
\end{figure}

In this model, the Sun is located at a distance $R_0=8.2\kpc$ \citep{Bland-Hawthorn2016} from the GC. The bar rotates at a pattern speed of $\Omega=39\kmskpc$ and is oriented at an angle of $\alpha=28\degree$ with the Sun$-$GC line of sight, consistent with the  measurements of \citet{Wegg2013} and \citet{Wegg2015} in the bulge and the long bar, respectively. The mass distribution of the model is such that the local circular velocity is $V(R_0)=243\kms$. Assuming a peculiar motion of the Sun with respect to the local circular orbit of $(U_0,V_0,W_0)=(11.1,12.24,7.25) \kms$ \citep{Schonrich2010}, the total tangential velocity of the Sun is $255\kms$ in approximate agreement with recent measurements \citep{Reid2004, Bovy2012,Schonrich2012,Reid2014}. The face-on projection of the model is shown in \autoref{fig:model}, together with the spatial coverage of the data described in the next subsection.

\begin{figure*}
  \centering
  \includegraphics{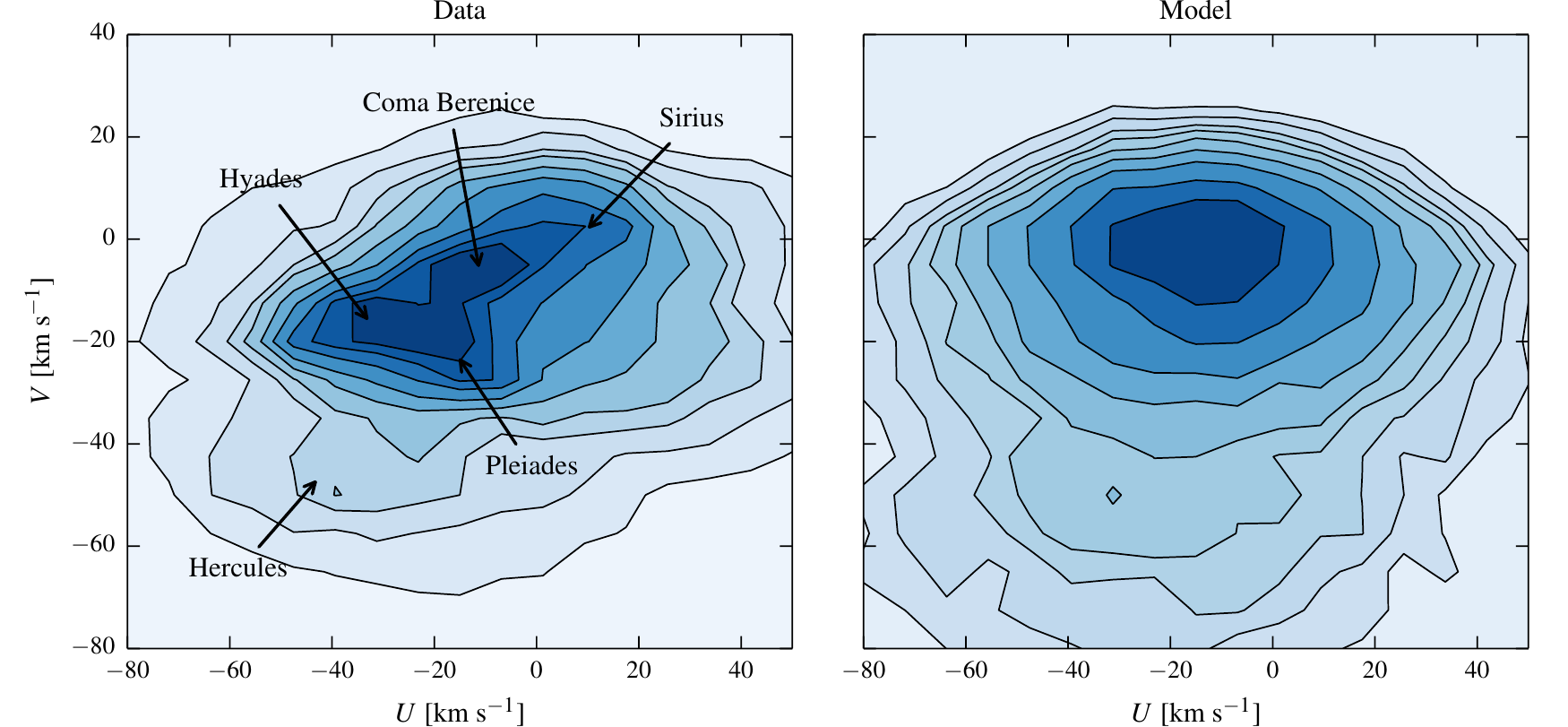}\\
  \caption{Velocity distribution in the SNd {\it U--V}-plane (relative to the Sun) for the data catalog (left) and the cross-matched model (right) within 300 pc from the Sun. The LSR is put at $U=-11.1\kms$, $V=-12.24\kms$. The contours contain, from inside outward, $10\%,20\%,30\%,40\%,50\%,60\%,80\%,90\%$ of the stars (left) or model particles (right).}
  \label{fig:model-data}
\end{figure*}

\subsection{6D phase-space data of SNd stars}
\label{section:data}

After one year of observation, the \gaia mission \citep{GaiaCollaboration2016a} recently released its first set of data \citep{GaiaCollaboration2016b} including the Tycho-Gaia Astrometric Solution (\tgas; \citealt{Michalik2015}), a large catalog of accurate astrometry, parallaxes, and proper motions for more than $2 \times10^6$ stars. Many of these \tgas stars have also been independently observed by spectroscopic surveys such as \rave (DR5; \citealt{Kunder2016}) and \lamost(star catalog from DR2; \citealt{Liu2014}), providing measurements for the radial velocities and metallicities of the stars.\footnote{The LAMOST velocities are corrected by $5.7\kms$ following \citet{Tian2015}} Thus, by cross-matching \tgas with \rave and/or \lamost, we construct a catalog  of 6D phase-space positions and metallicities of stars in the SNd. We limit ourselves to stars within $600\pc$ from the Sun with parallax errors smaller than $20\%$. For stars that have been observed spectroscopically multiple times (common stars between \rave and \lamost, or stars observed multiple times in \rave), we use as the radial velocity and metallicity the weighted mean of the various measurements available. Our final catalog  has 153,767 stars covering the SNd well, as displayed in \autoref{fig:model}. Finally, the R.A., decl, parallax, proper motions on the sky, and radial velocity of the stars are converted to Cartesian Galactic phase-space positions and velocities using \texttt{galpy} python tools \citep{Bovy2015}. 

\subsection{Cross-Matching Model and Data}
\label{section:matchingM-D}

Despite the increased resolution of the disk resulting from the resampling in \autoref{section:model}, the number of particles is too low to represent the velocity distribution in the SNd with sufficiently low noise. Therefore, we created a catalog  of $2\times10^6$ particles by integrating the model in its frozen potential and recording all particles within $600\pc$ from the Sun every $20\Myr$, the average time for particles to cross the SNd. This produces a large, time-averaged catalog and requires an integration time of $2\Gyr$ given our resolution.

In order to approximate the selection function of the data, we construct a cross-matched catalog, selecting from the particle catalog the nearest particle in space to each star from the data catalog. If this results in a duplicate selection, the next unselected particle is used. Henceforth, we will employ our cross-matched catalog where the number of particles is equal to the number of stars in our data catalog.

\section{Hercules Stream in the model}
\label{section:hercules}

In our model, the bar pattern speed is $39\kmskpc$, implying that the CR and OLR radius are at $6\kpc$ and $10.5\kpc$, respectively. Can a model with this value of pattern speed cause a bimodal velocity distribution? Which kind of orbits could generate a bimodality in the {\it U--V}-plane? In this section, we answer these questions and consider how the Hercules stream changes with Galactic radius and metallicity. 

\begin{figure*}
  \centering
  \includegraphics{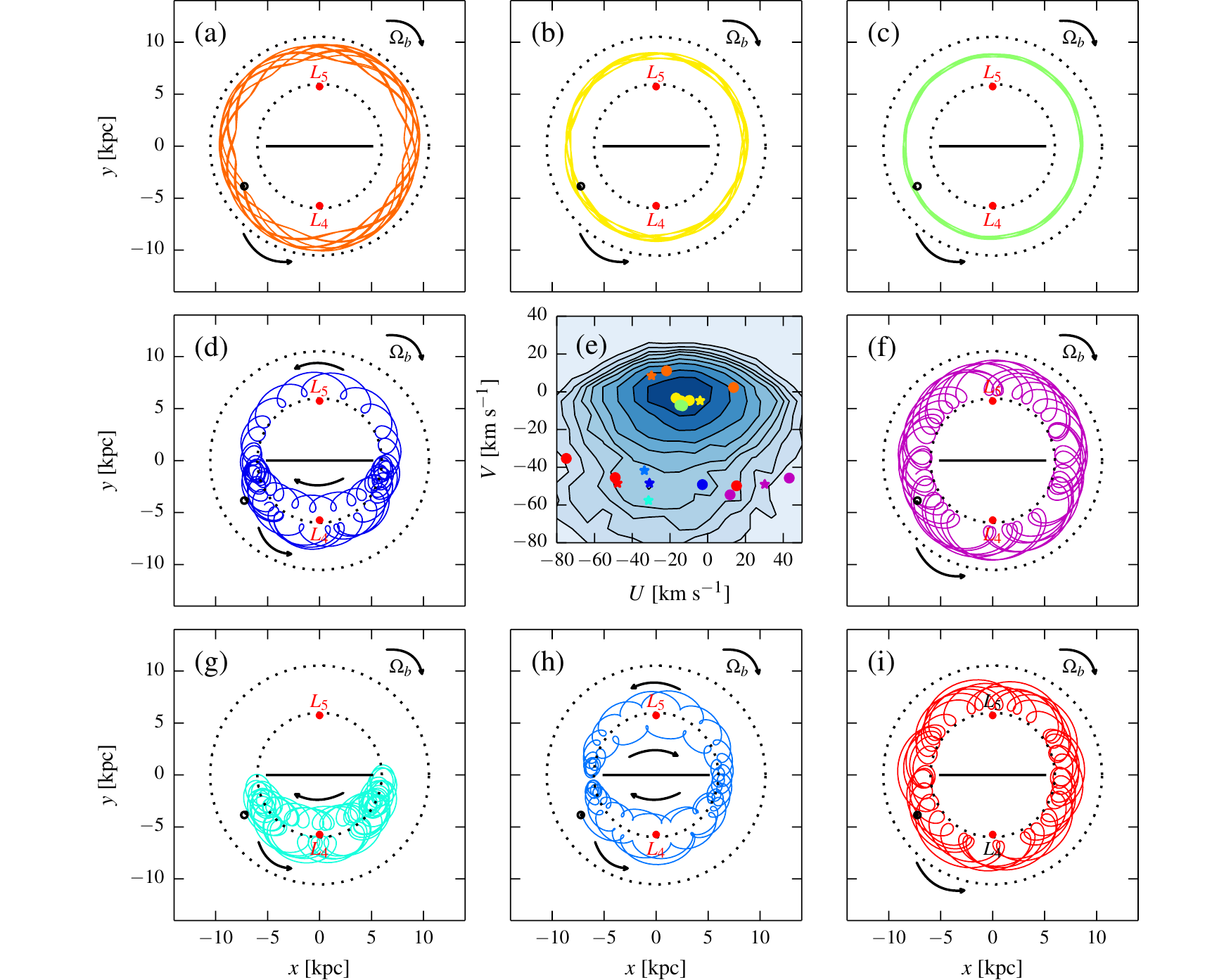}\\
  \caption{((a)-(c)) Representative orbits in the frame corotating with the bar for the main component. ((d),(g),(h)): Hercules stream orbits in the frame corotating with the bar. (e) Velocity distribution of the model within $300\pc$ from the Sun. The initial {\it U--V}-position where a particle is selected to construct its orbit is marked as a star, and the dots with the same color show  the {\it U--V}-positions where that particle again crosses the SNd during its orbit. Each color represents a different orbit. (d) An orbit near the Hercules density peak. (f) An orbit with opposite sign of $U$ with respect to (d). (i) Similar to (d), but at $U\simeq-50\kms$. The inner and outer dotted circles show the locations of the CR and OLR, respectively, and the red dots mark the positions of $L_4$ and $L_5$. The horizontal solid line shows the extension of the bar. The arrows along the orbits indicate the direction of motion in the corotating frame, while the arrows in the top right corner indicate the rotation of the bar in the inertial frame. }
  \label{fig:orbits}
\end{figure*}

\subsection{Comparison between Model and Data}
\label{sec:comparison}

In \autoref{fig:model-data}, we compare the SNd velocity distribution ({\it U--V}-plane) in the data (left) and in the model (right) within 300 pc from the Sun. We can see that the model naturally shows an excess of particles at $U\sim-30\kms$, $V\sim-50\kms$ in good agreement with Hercules. We note that the SNd velocity distribution has not been fitted but is a prediction.

In the observed {\it U--V}-plot, we can clearly identify several maxima in the main velocity component, including the Sirius and Pleiades--Hyades moving groups. These moving groups have been linked to a dynamical origin \citep{Famaey2008}  induced by either the bar  \citep{Minchev2010},
spiral arms  \citep{Simone2004,Quillen2005,Antoja2011}, or a mix of both \citep{Chakrabarty2007,Antoja2009}.

The main component in the model does not contain these maxima; however, this may be because this model does not include spiral arms.  The average $V$ velocity of these model stars is also roughly $10\kms$ larger than the circular velocity, due to streaming velocities induced in the bar. We estimated the radial velocity gradient in the model, $\vRR = C+K \approx-3.2 \pm 0.8\;\kmskpc$, and the Oort constants $C\approx-2.9\;\pm0.6 \kmskpc$ and $K\approx-0.3\; \pm0.6\kmskpc$, using local binned mean velocities and their gradients. $\vRR$ is at the lower end of measurements from \rave \citep{Siebert2011} and smaller than in \gaia DR1, $C+K\approx-6.6 \kmskpc$ \citep{Bovy2016}.

 To investigate further how these features and measurements could be explained and fitted in a more elaborate model is outside the scope of this Letter. Here, we concentrate on showing that the Hercules stream is a consequence of the bar and on understanding the orbits that are at its origin.

\begin{figure*}
  \centering
  \includegraphics{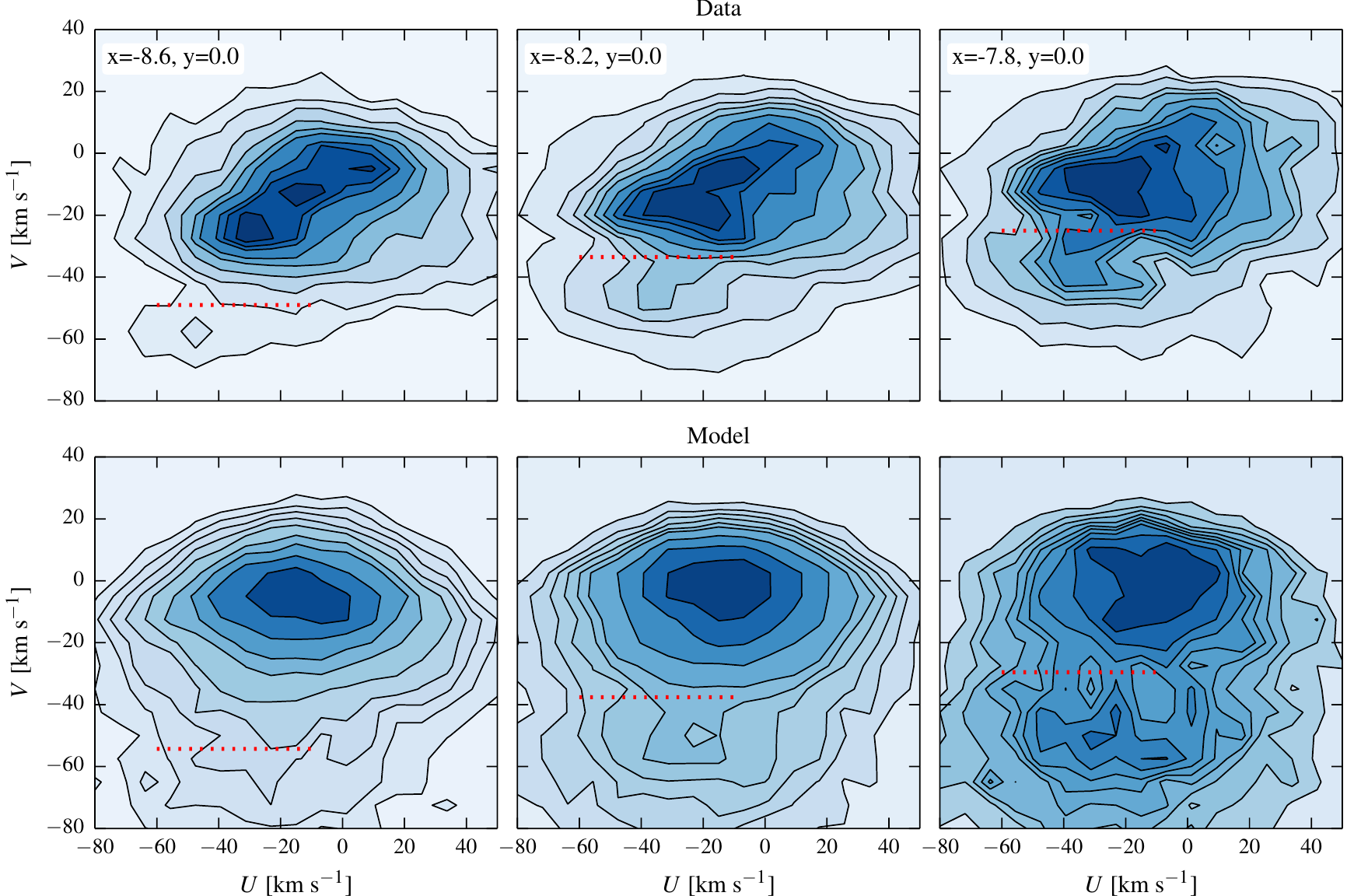}\\
  \caption{{\it U--V}-velocity distribution as function of Galactocentric distance, comparing data (top) and model (bottom). Each panel shows the velocity distribution in a square of $400\pc$, moving inward toward the GC from left to right. The center of each square in Cartesian coordinates is indicated in the top left corner of the data panels. Horizontal red dotted lines indicate minima or saddle points in the $V$-distribution integrated over the range $U=[-60, -10] \kms$ where Hercules is most prominent. The contours are as in \autoref{fig:model-data}. }
  \label{fig:following}
\end{figure*}

\subsection{Which orbits cause the bimodality in the {\it U--V}-plane?}
\label{section:orbits}

We have integrated for $5\Gyr$ in the rotating frozen potential a subset of the model particles within $300\pc$ of the Sun. \autoref{fig:orbits} shows the {\it U--V}-distribution in the SNd (\autoref{fig:orbits}(e)) from which we select orbits in the main component and in Hercules. The position in the {\it U--V}-plane where an orbit was started (at its coeval $x,y$ position) is represented by a star. Orbits in the main component (\autoref{fig:orbits}(a)-(c)) never cross the CR radius, and in general they have variable-sized epicycle motions around their guiding center. On the other hand, orbits in Hercules (\autoref{fig:orbits}(d), (g), (h)) have energies allowing them to cross the CR radius. In this region a class of orbits appears around the Lagrange points $L_4$ and $L_5$ on the minor axis of the bar that can be stable (the Lagrange points aligned with the bar,  $L_1$ and $L_2$, are unstable). \autoref{fig:orbits}(d) shows an orbit selected near the peak Hercules density. Note that the orbit with opposite sign of $U$ (\autoref{fig:orbits}(f)) is radially extended but remains outside CR, as does the orbit in \autoref{fig:orbits}(i). These orbits provide a constant background in the {\it U--V}-plot at negative $V$.

The Lagrange orbits librating around $L_4$ and $L_5$ have a banana-like shape parallel to the bar \citep{Contopoulos1989a}. In Hercules, we found mainly three kinds of Lagrange orbits: pure banana-like orbits orbiting around $L_4$, orbits that circulate $L_4$ and $L_5$, and orbits that move around CR for a while but then they get trapped by the $L_4$ or $L_5$ Lagrange point. The latter two classes are probably chaotic orbits. All three classes visit the SNd, but only rarely as their $V$-velocity in the corotating frame is small. They cross the SNd with $U<0$ moving outward, in contrast to a perturbed axisymmetric distribution function just outside corotation in slow bars  \citep{ Monari2017a}, which has a symmetric feature in $U$ with $V<0.$

\subsection{Hercules at Different Galactic radii}
\label{section:following-hercules}

With the entire sample of stars that we have in our data catalog  around the SNd within $600\pc$ from the Sun, we can make cuts in galactocentric distance in order to determine where the Hercules stream can be detected. We make three square bins of $400\pc$, where the center of each bin, in Cartesian coordinates with respect to the GC, in kpc, is $(x,y)=(-8.6,0), (-8.2,0),$ and $(-7.8,0)$. Because the Hercules stream is formed by stars that travel from the inner Galaxy to the SNd, we should expect it to be most prominent in the innermost bin.

In \autoref{fig:following}, we show the velocity distribution in the three square bins for the data (top) and the model (bottom). In the data, Hercules is indeed most prominent in the inner bin, is still clearly present in the middle bin centered at the Sun's position, and then is almost absent in the outer bin only a few 100 pc
outside the solar radius.  The same trend is present in our model: Hercules is very clear in the innermost bin, and it disappears in the outermost bin. Therefore, our new interpretation is consistent with the observed variation of the strength of Hercules at different Galactocentric radii. \citet{Bovy2010a}, \citet{Antoja2014} and \citet{Monari2017} considered the position of the ``gap'' between the main component and Hercules as function of galactocentric distance. For all bins in \autoref{fig:following}, we estimated the position of the gap as function of distance, by looking for a minimum or saddle point in the $V$-distribution, $V_g$ \citep[similar to][]{Monari2017}. $V_g$ decreases outward in both the data and the model, as indicated in \autoref{fig:following} but it is clear from the {\it U--V}-diagrams that further work is required to match quantitatively the Hercules peak inward from the Sun.

\subsection{The {\it U--V}-plane in Different metallicity ranges}
\label{section:metallicity}

In our explanation, the stars in the Hercules stream come from around the bar's CR radius, implying that these stars should on average be older and more metal-rich than the SNd stars. This has indeed been observed by \citet{Liu2016} for the \lamost stars, even though as first shown by \citet{Raboud1998}, the Hercules stars have a wide metallicity distribution.

\begin{figure}
  \centering
  \includegraphics{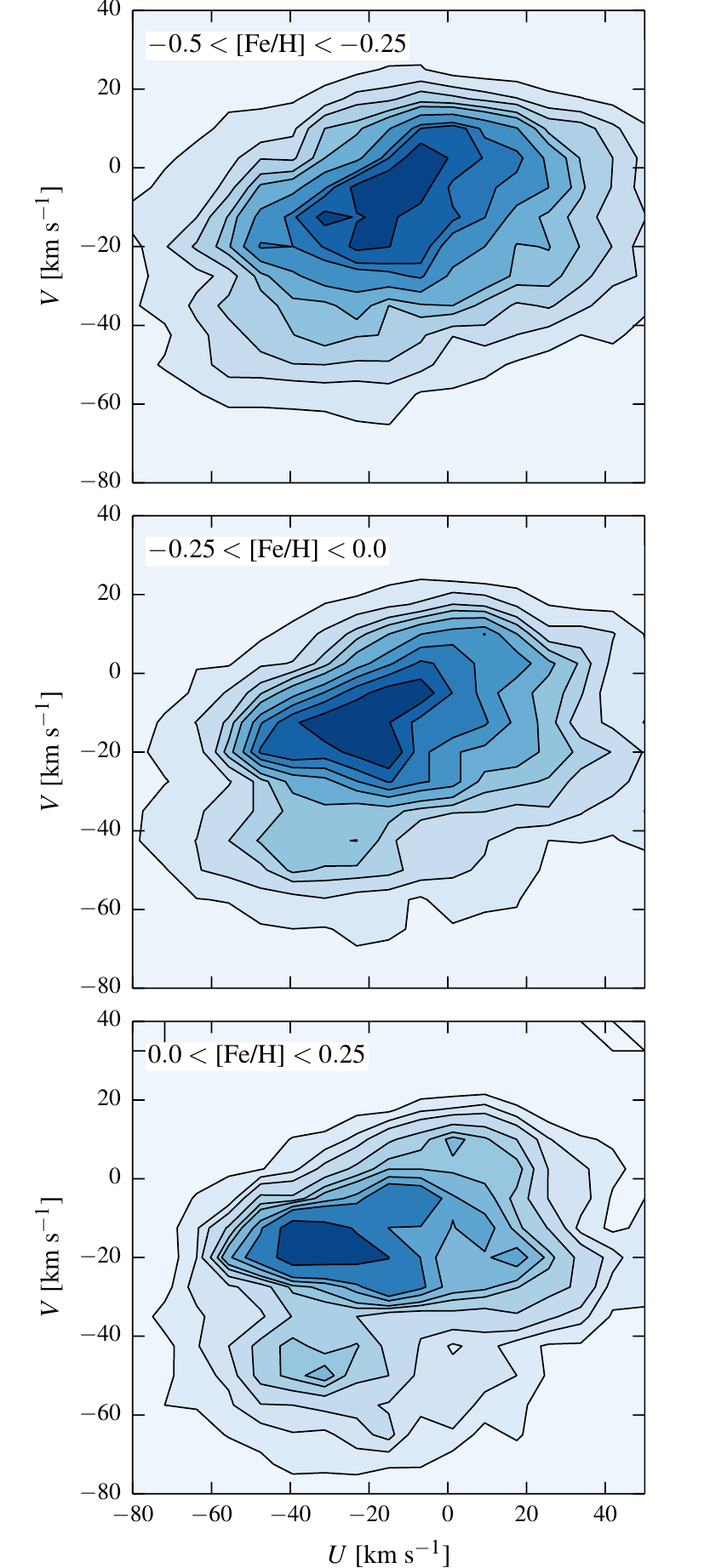}\\
  \caption{Velocity distribution of the \tgas+ \rave data in different metallicity intervals. The contours are as in \autoref{fig:model-data}.}
  \label{fig:metallicity}
\end{figure}

In \autoref{fig:metallicity}, we show the {\it U--V}-velocity distribution for different metallicity intervals, $-0.5<$[Fe/H]$<-0.25$ (top), $-0.25<$[Fe/H]$<0.0$ (middle), and $0.0<$[Fe/H]$<0.25$ (bottom) for the  RAVE+TGAS sample. We did not use our entire catalog  because the metallicity scale of the \rave and \lamost surveys is not the same. \autoref{fig:metallicity} shows that Hercules is less conspicuous in the most metal-poor stars, but more prominent in the most metal-rich stars. Again this is consistent with the idea that the Hercules stars come from the inner Galaxy.  A similar effect is seen in the main {\it U--V}-component where the Hyades and Pleiades groups dominate the most metal-rich stars, suggesting that also these stars mainly originate inside the solar radius.

\section{Summary and discussion}
\label{section:summary}

The {\it U--V}-velocity distribution observed in the SNd is complex, and explaining every detail is not an easy
task. This letter is devoted to giving a new explanation for the dynamical origin of the Hercules stream.
The standard explanation for Hercules is that it is generated by orbits scattered by the
OLR. However, this explanation requires the OLR to be close to the Sun \citep{Dehnen2000, Antoja2014,Monari2017}, which is incompatible with recent measurements of a lower pattern speed \citep{Portail2017} and long bar \citep{Wegg2015}. For these low pattern speeds, the OLR is placed significantly beyond the SNd and cannot be responsible for the origin of Hercules.

We have analyzed the velocity distribution in the SNd using the best M2M particle model of the Galaxy of
\citet{Portail2017} fitted to a variety of inner Galaxy data.  This model has a bar pattern speed of $39 \kmskpc$. Without fitting the SNd velocity distribution, we find that this model naturally contains a secondary peak in the {\it U--V}-plane, which is consistent with the Hercules stream as seen in the TGAS-RAVE-LAMOST data. In
the model, Hercules is mainly made of stars orbiting the Lagrange points near the bar's CR radius that have the right energies to visit the SNd. Some of these orbits circulate around $L_4$ only, but others go also to $L_5$.

Our new model naturally predicts that the Hercules stream is more prominent inward toward the GC and almost absent beyond a few 100 pc outward from the Sun, as seen in the TGAS-RAVE-LAMOST data. This trend may also be compatible with the OLR model \citep{Bovy2010a,Antoja2014,Monari2017}. However, only the explanation advocated here is also consistent with the dynamics of the bar and bulge region. Our explanation also very plausibly describes the fact that the Hercules stars are more prominent in old and metal-rich stars, as seen in \citet{Liu2016} and in \autoref{fig:metallicity}. 

We do not yet understand how the substructure in the main distribution of SNd velocities arises in the context of the longer and slower bar and what role the Galactic spiral arms play in this. Better models of the SNd distribution function are also required to test whether the position in velocity of Hercules can change in space correctly. Understanding these points will be greatly aided by $Gaia$ DR2, which will allow us to make {\it U--V} diagrams out to several kiloparsecs in the disk and provide definitive answers on the nature of Hercules and indeed all the dynamical streams in the disk.

\acknowledgments

We thank Martin Smith for pointing out the paper by \citet{Tian2015} and Jo Bovy for making available his tool dealing with {\it Gaia} data. A.P.V. is grateful for a Conacyt postdoctoral fellowship and an MPE visiting fellowship. We also thank the anonymous referee for their careful reading and constructive comments. 
This work has made use of data from the European Space Agency {\it Gaia} mission (\url{http://www.cosmos.esa.int/gaia}), processed by the {\it Gaia} Data Processing and Analysis Consortium (DPAC), of
RAVE data (\url{www.rave-survey.org}), and of data from the LAMOST survey of the Guoshoujing Telescope built by the Chinese Academy of Sciences (\url{www.lamost.org}).

\end{document}